\documentclass[sigconf]{acmart}
\usepackage{soul}
\usepackage{todonotes}
\AtBeginDocument{%
  \providecommand\BibTeX{{%
    \normalfont B\kern-0.5em{\scshape i\kern-0.25em b}\kern-0.8em\TeX}}}

\setcopyright{acmcopyright}
\copyrightyear{2019}
\acmYear{2019}
\acmDOI{}

\copyrightyear{2019}
\acmYear{2019}
\acmConference[SUI '19]{Symposium on Spatial User Interaction}{October 19--20,
	2019}{New Orleans, LA, USA}
\acmBooktitle{Symposium on Spatial User Interaction (SUI '19), October 19--20, 2019, New
	Orleans, LA, USA}
\acmPrice{15.00}
\acmDOI{10.1145/3357251.3357588}
\acmISBN{978-1-4503-6975-6/19/10}

\begin{document}

\title{Effects of Depth Layer Switching between an Optical See-Through Head-Mounted Display and a Body-Proximate Display}

\author{Anna Eiberger}
\affiliation{%
  \institution{University of Passau}
}
\email{eiberg01@stud.uni-passau.de}

\author{Per Ola Kristensson}
\affiliation{%
	\institution{Cambridge University}
}
\email{pok21@cam.ac.uk}

\author{Susanne Mayr}
\affiliation{%
	\institution{University of Passau}
}
\email{susanne.mayr@uni-passau.de}

\author{Matthias Kranz}
\affiliation{%
	\institution{University of Passau}
}
\email{matthias.kranz@uni-passau.de}

\author{Jens Grubert}
\affiliation{%
	\institution{Coburg University of Applied Sciences and Arts}
}
\email{jens.grubert@hs-coburg.de}
\renewcommand{\shortauthors}{anonymous}

\begin{abstract}
  Optical see-through head-mounted displays (OST HMDs) typically display virtual content at a fixed focal distance while users need to integrate this information with real-world information at different depth layers. This problem is pronounced in body-proximate multi-display systems, such as when an OST HMD is combined with a smartphone or smartwatch. While such joint systems open up a new design space, they also reduce  users' ability to integrate visual information. We quantify this cost by presenting the results of an experiment ($n$=24) that evaluates human performance in a visual search task across an OST HMD and a body-proximate display at 30 cm. The results reveal that task completion time increases significantly by approximately 50\% and the error rate increases significantly by approximately 100\% compared to visual search on a single depth layer. These results highlight a design trade-off when designing joint OST HMD-body proximate display systems.
\end{abstract}



\begin{teaserfigure}
 \centering
 \includegraphics[width=0.8\columnwidth]{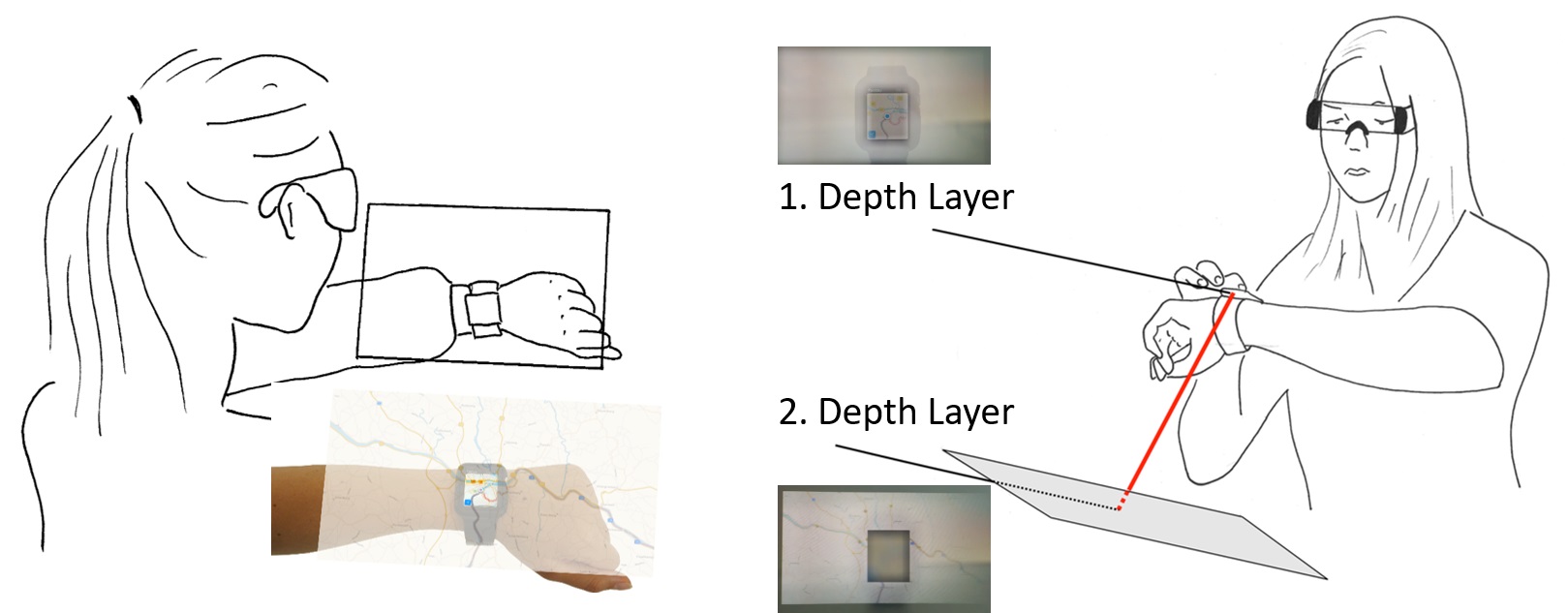}
 \caption{Left: Vision of extending the field of view of body-proximate displays such as a smartwatch through an optical see-through head-mounted display with visual information presented at a uniform depth layer. Right: Visualization of the two depth layers (smartwatch and focal distance of the head-mounted  display), which leads to the need of accomodation and vergence changes to integrate information accross both displays. }
 \label{fig:vision}
\end{teaserfigure}

\maketitle

\section{Introduction}
\begin{figure*}[htb!]
	\centering 
	\includegraphics[width=1.6\columnwidth]{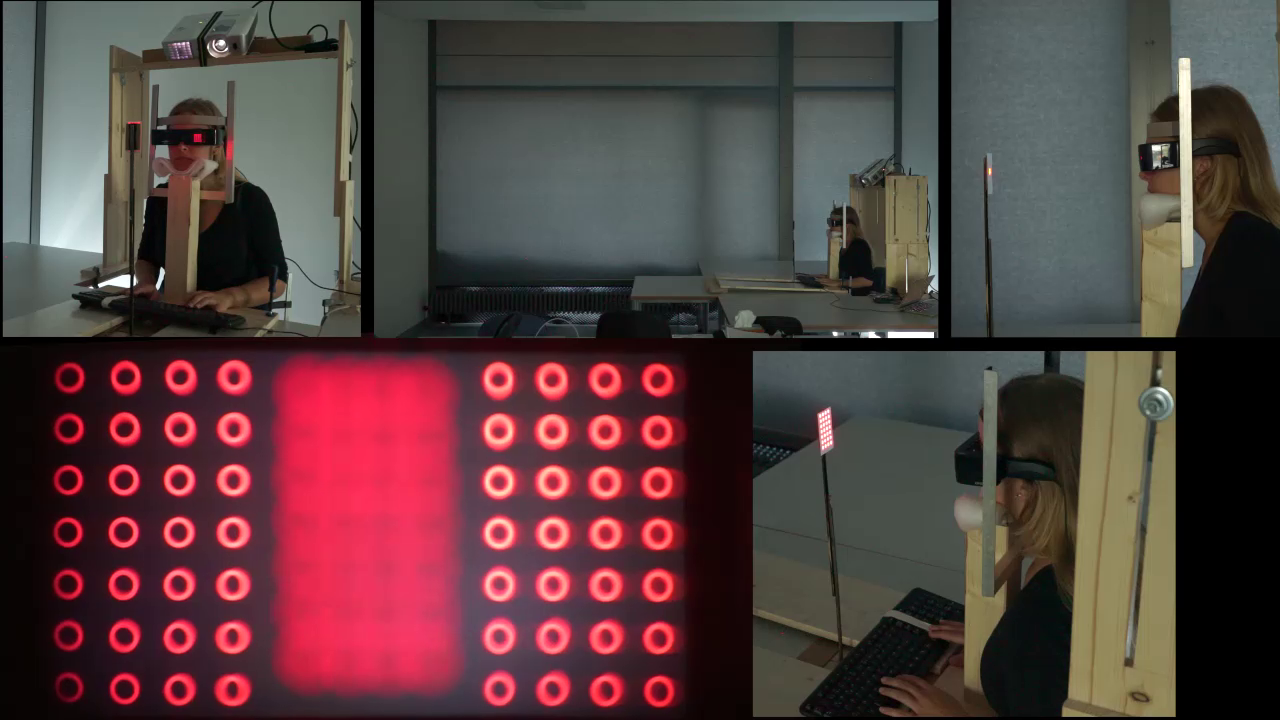}
	\caption{Multiple views on a participant in condition \textsc{HMD+Projector}. Top left: Front view. Top middle: side view of the experiment room. Top right: close-up side view with projection area on the left. Bottom left: view through the HMD with focus on the HMD plane (hence, the center four columns appear blurred)}. Please note, that in the actual first person view of the user, she would see double images of the plane out of focus due to stereopsis). Bottom right: Back view with projection area.
	\label{fig:conditiondualfocus}
\end{figure*}

Optical see-through head-mounted displays (OST HMDs) are on the verge of becoming mainstream computing platforms, and certain devices, such as the Microsoft HoloLens, are already in-use in various industries, such as construction engineering.

Body-proximate displays, such as smartwatches and smartphones have the potential to augment OST HMDs \cite{grubert2015multifi, normand2018enlarging} and provide two primary benefits. First, they can extend the available display space and thereby provide new opportunities for distributing information to the user (see also Figure \ref{fig:vision}, left). Second, they can introduce a secondary capacitive touchscreen input channel to the OST HMD system. As pure OST HMD input is inherently noisy and error-prone due to a reliance on voice, eye-tracking, head-tracking or gesture recognition, such a secondary input channel is valuable as it provides relatively certain low-noise information about users' intentions. For instance, a joint OST HMD-smartwatch system was recently studied with respect to indirect control of the cursor on the OST-HMD via the smartwatch \cite{wolf2018performance}. 

Other potential applications include distributing notifications or warnings across a phone and an OST HMD, showing context-relevant information about the visual scene depicted in the OST HMD on the phone, or augmenting OST HMD interaction with a physical input/output device, for instance, by allowing users to move their phone in front of their face and observe a dynamic lens of the visual scene which can be modulated by the user interacting directly with the phone.

However, while body-proximate displays may provide these and possibly other benefits to OST HMD systems, the implications of using such joint systems are currently underexplored. In general, while distributing information across a body-proximate display and an OST HMD can open up new user interface solutions, such distribution of information is likely to negatively affect users' ability to integrate information (see Figure \ref{fig:vision}, right). 
Hence, it is neccessary to quantify the cost of jointly processing visual information across OST HMDs and body-proximate displays.

With the exception of Magic Leap Creator's Edition, commercial OST HMDs present virtual information at a single focus distance. Vari-focal HMDs (e.g., \cite{dunn2017wide, akcsit2017near, wilson2018high}) are also still not widely commercially available.

Prior work \cite{huckauf2010perceptual, gabbard2018effects} indicates that if information in the physical scene has to be integrated with virtual information, the user's visual and cognitive load increases. However, this prior research has mostly focused on users integrating virtual information on an OST HMD display with information in the physical scene in the medium or far visual field, such as visual search tasks or depth perception judgment tasks. Work focusing on OST HMD information integration in the near field are rare, with a few exceptions, such as work investigating depth-perception \cite{singh2018effect} and object localization \cite{ellis1998localization}.

In this paper, we complement prior research efforts by investigating the joint performance of an OST HMD coupled with a body-proximate display. We report of an experiment comparing reaction time, error rate, workload and degree of simulator sickness induced in a visual search task carried out either solely in the depth field of the OST HMD or distributed across the depth field of the OST HMD and the body proximate display. We find that the distribution of the visual search task results in a significant (approximately 50\%) increase in reaction time and a significant (approximately 100\%) increase in error rate as well as in increases in workload, however, no significant difference in perceived simulator sickness. These results quantify the substantial information integration cost a joint OST HMD-body-proximate system incurs for users and suggest that there is a significant trade-off when using joint OST HMD and body-proximate display systems.

\section{Related Work}

The work builds on a large body of related work from the domains of perceptual psychology, augmented reality and multi-display systems. In the following, we highlight work from the domains body-proximate multi-display systems and perceptual issues in OST HMD.

\subsection{Perceptual Issues in OST HMDs}

Commercially available binocular OST HMDs display virtual content typically at a single focal distance. Various perceptual issues when using OST HMDs have been explored (c.f. \cite{kruijff2010perceptual} for an overview). Specifically, if content from the physical world at a different focal distance needs to be integrated with the virtual content, this can lead to the need of repeated accommodation changes as well as context switches, resulting in increasing visual and cognitive attention \cite{gabbard2018effects}. Relevant factors are differences in focal distance and vergence as well as the visual angle between physical and virtual content.

Depending on the expected distance of content in the physical world, different focal distances have been proposed for Head-up displays (e.g., 2m or optical infinity) with no concluding recommendations, yet \cite{weintraub1992human, wolffsohn1998effect}. Gabbard et al.  highlighted the need for multi-focal Augmented Reality (AR) displays \cite{gabbard2014behind} with several research protoypes beginning to emerge \cite{dunn2017wide, akcsit2017near, wilson2018high}. For current single focus HMDs, Oshima et al. \cite{oshima2016sharpview} and Cook et al. \cite{cook2018user} proposed and evaluated a system for adaptive sharpening HMD display content. Due to commercial availability, our study used a collimated off-the shelf OST HMD. 

There have been a number of studies, investigating the effects of processing information across OST HMDs and the physical world. 
Huckauf et al. \cite{huckauf2010perceptual} investigated the effects of context switching (i.e. switching of visual and cognitive attention between virtual and physical world information) between a monocular OST HMD and a CRT-monitor, both placed at the same focal distance of 61~cm. Their results indicate that there is a cost of context switching ca. 10\% between a CRT and an OST-HMD when both displays are at the same focus distance).

Gabbard et al. \cite{gabbard2018effects} also investigated the effects context switching, but in conjunction with focal distance switching (i.e. accomodation change). The investigated focal distances were 0.7~m, 2~m and 6~m. Their findings indicate that both focal distance and context switching result in significantly reduced performance in a reading task. Their work shows that both context and focal distance switching are important AR user interface design issues. Our study builds up on this prior work by investigating the joint performance of a (handheld or body-worn) display at typical viewing distance of a smartphone \cite{bababekova2011font} with an OST HMD.

Winterbottom et al. \cite{winterbottom2007depth} studied multi-focal AR displays and empirically determined suitable focus distances for in the context of a flight simulator. 

Within near-field reaching distances, Singh et al. \cite{singh2018effect} investigated the effects of focal distance, age and brightness when matching the depth between a real and virtual object using an haploscope within distances between 33.5-50~cm in a depth-matching task \cite{swan2015matching}. Their results indicate that collimated graphics (i.e. graphics shown at infinite focal distance) resulted in inferior performance compared to graphics shown at a nearer focal distance.

Ellis and Menges \cite{ellis1998localization} studied the effects of vergence, accommodation, observer age, viewing condition  and the presence of an occluding surface in the near-field. They found that accuracy is degraded by monocular viewing and an occluding surface. McCandless et al. [42] studied motion parallax and latency in monocular viewing and found reduced accuracy with increasing distance and latency. Singh et al. \cite{singh2010depth} found that an occluding surface has complex accuracy effects, and Rosa et al. \cite{rosa2016visuotactile} found that accuracy increased when using redundant tactile feedback.

Our study builds on this prior work and extends it to the domain of body-proximate multi-display systems.

\subsection{Body-Proximate Multi-Display Environments}

Combining mobile body-proximate displays into multi-display environments has been repeatedly investigated. However, the majority of previous works has studied the joint interaction between multiple handheld devices, such as smartphones or tablets, body-worn devices, such as smartwatches or pico-projectors. Only few works dealt with joint interaction between HMDs and smartwatches or smartphones.

Chen et al. studied joint interactions between a smartphone and a smartwatch \cite{chen2014duet}. R\"adle et al. \cite{radle2014huddlelamp} proposed a lamp-mounted spatial tracking system for interacting accross multiple tablets . Grubert et al. \cite{grubert2017headphones, grubert2017towards}  extended the idea to not use an instrumented environment, but the user's head as reference point instead.

Houben et al. presented a toolkit for prototyping smartwatch-centric cross-device interactions \cite{houben2015watchconnect}. A number  of different cross-device toolkits have been proposed for focusing on smartwatch, web, or Augmented Reality centric development aspects, e.g., \cite{chi2016enhancing, husmann2017investigating, nebeling2014xdkinect, schreiner2015connichiwa, seyed2015sod, speicher2018xd}.

Grubert et al. \cite{grubert2016challenges, quigley2015perceptual} presented a survey on challenges in mobile multi-display ecologies and highlighted potential perceptual issues in body-proximate multi-display systems, such as display contiguity \cite{cauchard2011visual, rashid2012factors}, visual attention \cite{rashid2012cost, neate2015mediating, holmes2012visual} or visual overload \cite{dostal2013subtle, neate2015designing} which was followed up by a recent survey of Brudy et al. \cite{brudy2019cross}.

\begin{figure*}[h!]
	\centering 
	\includegraphics[width=1.8\columnwidth]{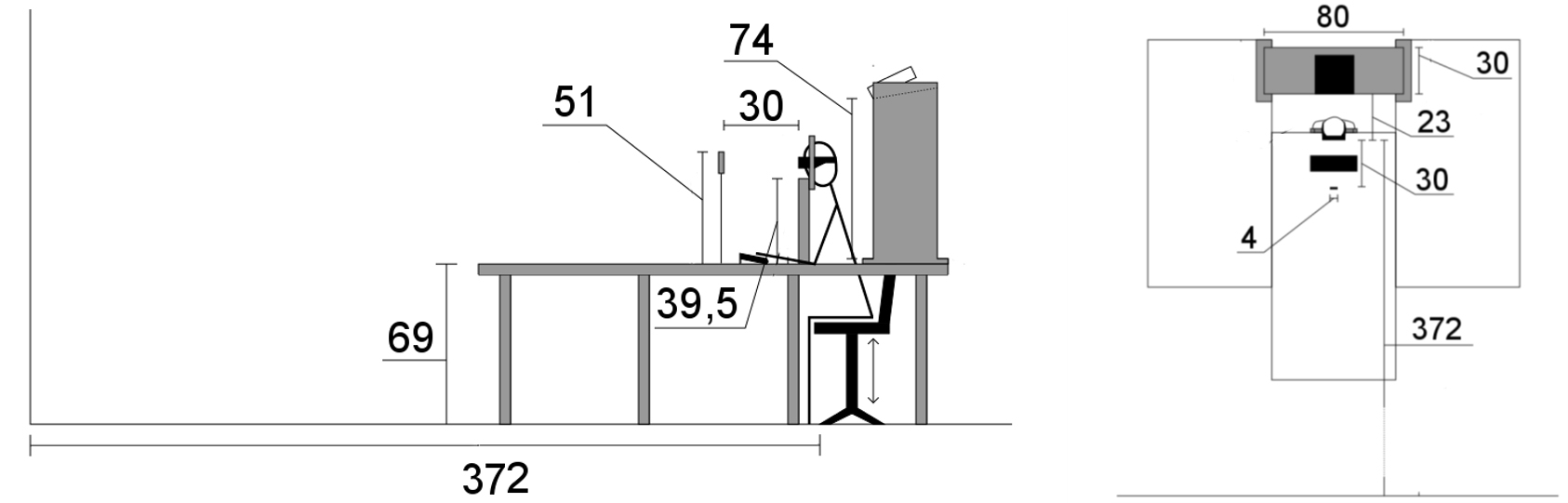}
	\caption{Side view (left) and top view (right) on the apparatus setup. All distances in cm.}
	\label{fig:apparatusconcept}
\end{figure*}

Regarding the use of OST HMDs in multi-display environments, Serrano et al. \cite{serrano2015gluey} investigated the joint use of OST HMDs within a desktop environment. Grubert et al. \cite{grubert2015multifi} proposed several interaction techniques between an OST HMD, a smartwatch and smartphone and conducted performance tests of the joint system, but did not compare it against a single device baseline. Budhirja et al. \cite{budhiraja2013using} proposed using a smartphone as indirect pointing device for an HMD, an idea late extended unobtrusive input devices \cite{dobbelstein2017pocketthumb}. Wolf et al. \cite{wolf2018performance} determined the performance envelope of in-air direct and smartwatch indirect control for OST HMDs.

Our work informs researchers, designers and developers of joint interactive systems between OST HMDs and body-proximate displays, such as smartwatches and smartphones about the costs of joint information presentation accross those devices.



\section{Experiment}

The experimental design is inspired by previous experiments on context switching and focal distance switching \cite{huckauf2010perceptual, gabbard2018effects} and extends those previous investigations to a combination of a OST HMD and a body-proximate display. The primary research question we are addressing is a quantification of how well users are able to integrate visual information jointly presented across an OST HMD and a display at a typical viewing distance of a smartphone \cite{bababekova2011font}.

To this end, we asked participants to conduct a visual search task either on an HMD alone or across the HMD and a display at a closer focus distance.


\begin{figure*}[th!]
	\centering 
	\includegraphics[width=2.0\columnwidth]{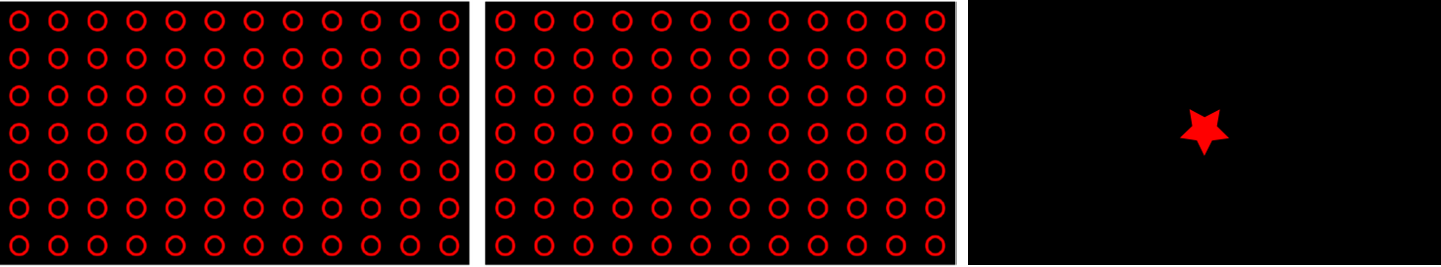}
	\caption{Left: stimulus image with no target present. Middle: stimulus image with target present in row 5, column 7. Right: secondary task with star pointing downwards.}
	\label{fig:focustask}
\end{figure*}

\subsection{Design}

The experiment used a within-subjects design with one independent variable \textsc{DepthLayers}. It had two levels: In the \textsc{HMD+Projector} condition the task was distributed across the HMD (in stereo) and a second screen illuminated by a projector in front of the user. In the \textsc{HMD} condition, the task was conducted at the single HMD screen in stereo mode. We decided against a possible third condition, conducting the task using the projector only at near distance as a pilot study indicated no significant differences between projector only and \textsc{HMD}.

\subsection{Participants}
Twenty-four volunteers participated in the experiment (12 male, 12 female, mean age 24.7 years, sd = 4.3). All but one participants exhibited normal or corrected to normal vision using corrective glasses or contact lenses. One participant exhibited 20/33 (6/10, logMAR: 0.22) visual acuity with contact lenses but could conduct the study without problems.  

\begin{figure}[t!]
	\centering 
	\includegraphics[width=\columnwidth]{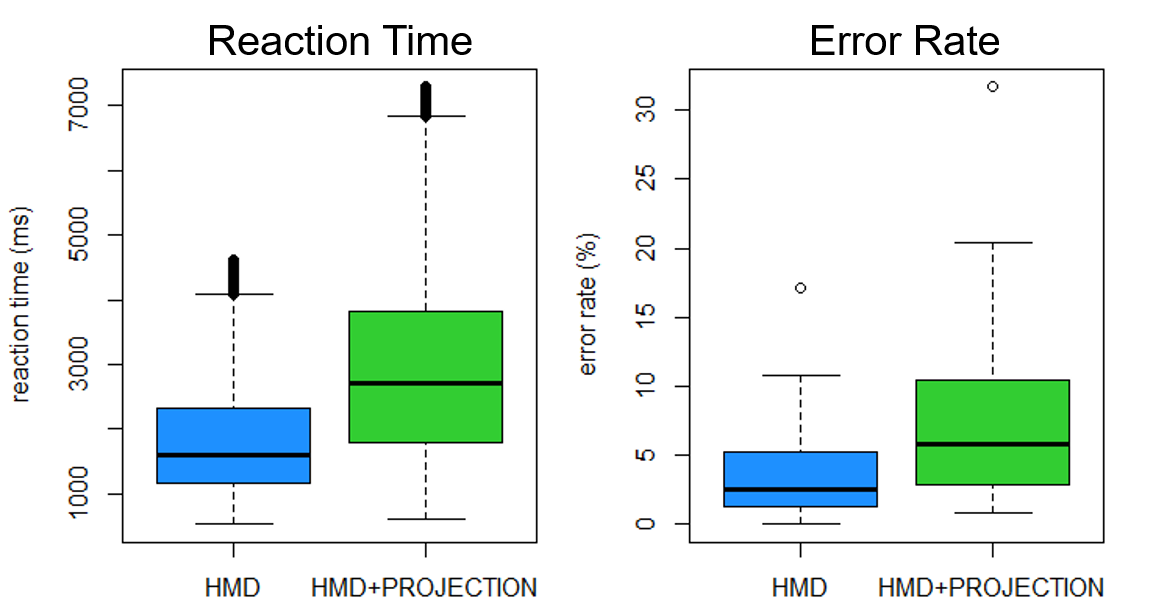}
	\caption{Left: reaction time in ms. Right: error rate.}
	\label{fig:resultse5}
\end{figure}

\begin{figure}[tbh]
	\centering 
	\includegraphics[width=\columnwidth]{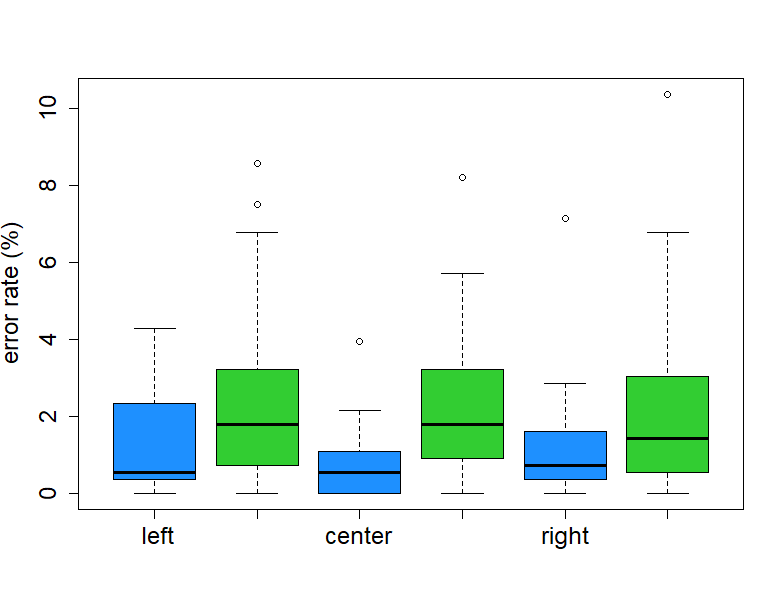}
	\caption{Error rate split between left, center and right zones, with each zone encompassing 4 columns. In the condition \textsc{HMD+Projector} (green) the left and right zone correspond to the columns presented on the HMD and the center zone corresponds to the columns presented on the projector. Condition \textsc{HMD} is depicted in blue. No significant differences where indicated between the zones}
	\label{fig:results-zones}
\end{figure}

\subsection{Apparatus and Materials}

The apparatus is shown in Figures \ref{fig:apparatusconcept} and \ref{fig:conditiondualfocus}.  A Benq W1070 Full HD projector was used for projection on 4 $\times$ 7.5 cm sheet of cardboard in 30 cm distance of the user. This distance corresponds to a typical smartphone and smartwatch viewing distance \cite{bababekova2011font, blascheck2018preparing}.
The projector was mounted on a stand above the user's head. The HMD used was an Epson Moverio BT-100 (horizontal field of view: 30.3\textdegree vertical field of view: 11.4\textdegree). The HMD displayed content at infinite focus and the default vergence distance of 370 cm, which corresponded to a wall at the viewing direction of participants. The field of view of the HMD was too small to set the vergence distance at the same distance as the projector. The head of the user was fixated on a chin rest (two chin rests were used to accommodate for different user heights). Colors, brightness and resolution of the HMD and the projector were matched as close as possible. A node.js server was used to synchronize the image presentation between the projector and the HMD. There was no noticeable delay between projector and HMD image presentation.

The use of other displays, such as a smartwatch or smartphone panel display or other HMDs (as well as resolutions, contrasts,  slight variations in distance) likely would lead to variations in the absolute values of the results to be reported. In a pre-experiment, we considered a panel display (Samsung S8+ smartphone) as body-proximate display, which indicated comparable results to the projector. We ultimately chose a projector as the smartphone (curved) bezel would have interfered with the display of the adjacent symbols on the projector. Also, we are not aware of evidence suggesting that the significance or scale of results would substantially change by interchanging a projector display with a panel display. 

A physical keyboard, placed on a table at the users hands, was used to get user feedback for both tasks.

For the primary task, a matrix of 84 circular symbols arranged in 7 rows by 12 columns was presented, see Figure \ref{fig:focustask}, left and middle. 

The target symbol "0" (number zero) had a width of 0.66\textdegree~($40'$) and height of 1.00\textdegree~($60'$).
The distractor symbol "O" (letter O) had a width of 0.85\textdegree~($51'$) and height of 0.94\textdegree~($56'$). 

In condition \textsc{HMD}, all symbols were displayed on the HMD. In the condition \textsc{HMD+Projector} the middle 4 columns were presented on the display at 30 cm distance. The target would show up with a likelihood of 50\% and its location on the grid was uniformly distributed in both conditions.

In condition \textsc{HMD+Projector}, if the target would show up, then it would appear in 33.3\% of all cases on the 4 left, 4 middle or 4 right coloumns (i.e. 66.6\% on the HMD and in 33.3\% on the front projection). The left arrow key should be pressed if the target was present, otherwise the right arrow key. After pressing a key an audio signal would indicate whether the decision was correct or not. The participants had the chance of getting used to the mapping in a trial session. 


The secondary task was conducted between images of the primary task, i.e. in alternating order. Here, a star shape would show up, see Figure \ref{fig:focustask}, right. The participant was asked to decide if a star prong was pointing upwards (B) or downwards (space bar). In condition \textsc{HMD} the star was shown on the HMD and in condition \textsc{HMD+Projector} it was shown on the front projection.

\subsection{Task and Procedure}

The primary task was to identify whether a target symbol (``0'', the number zero) was present among distractors (``O'', the letter O). In a secondary, alternating task, participants had to decide whether a star was pointing upwards or downwards. The purpose of the secondary task was solely to erase the previous picture from the primary task from the participant's visual memory. Hence, we exclude the reporting of the results for the secondary task.

Each participant filled out a demographic questionnaire and their visual acuity was determined using the Freiburg visual acuity test \cite{bach1996freiburg} to ensure that the displayed symbols could be recognized. All participants were found to be suitable for inclusion into the experiment.


For each condition, the participant was shown 280 separate symbol matrices executed in 4 separate blocks of 70 images each in the primary task, resulting in a total of 560 task executions per participant. The order of the starting condition was counterbalanced across all participants. After each condition the participants filled out the unweighted NASA TLX \cite{hart1988development} questionnaire and the Simulator Sickness Questionnaire (SSQ) \cite{kennedy1993simulator}.  

The experiment was carried out in a single ca. 60-minute session structured as a 20-minute introduction and briefing phase, a 30-minute testing phase (ca. 10 minutes for condition \textsc{HMD}, ca. 15 minutes for condition \textsc{HMD+PROJECTOR} + five-minute breaks including questionnaires), and 10 minutes for final questionnaires, interview and debriefing. Also, the participants were allowed to rest between individual blocks but did not make use of this option.

\subsection{Results}

Statistical significance tests for reaction time were carried out using paired, two-tailed t-tests with a significance level of $\alpha = 0.05$. Gender-specific differences for reaction time were tested for using indeptendent t-tests. The analysis were checked for appropriateness against the dataset, log transformed data was used for reaction time. Subjective feedback and error rate were analyzed with non-parametric Wilcoxon signed-rank tests. Gender-specific differences were tested for using Mann-Whitney U tests.

As the purpose of the secondary task was to erase the previous picture from the primary task from the participant's visual memory, we report on the resulty of the primary task only. Next, we report on reaction time, error rate (i.e. the number of false user decisions divided by the total number of stimuli images), workload (as measured by NASA TLX \cite{hart1988development}) and simulator sickness (as measured by the simulator sickness questionaire SSQ  \cite{kennedy1993simulator}). We further report on the search strategies employed by participants.

\subsubsection{Reaction Time} 

The mean reaction time for \textsc{HMD} was 2171 ms (sd = 846) and for \textsc{HMD+PROJECTOR} 3233 ms (sd = 1217), see Figure \ref{fig:resultse5}, left. A two-tailed paired sample t-test on log transformed data (to reduce skewness of the data) indicated a significant difference in reaction time ($t(23) = -5.58, p < 0.001, Cohen's~d = 1.14$). 
Also, no block-wise or gender-specific differences within conditions for reaction time were found. 

In other words, the reaction time using \textsc{HMD+PROJECTOR} significantly increased about 50\% compared to only \textsc{HMD}. 

\subsubsection{Error Rate}
The mean error rate for \textsc{HMD} was 3.84\% (sd = 3.95) and for \textsc{HMD+Projector} 8.30 (sd = 7.68), see Figure \ref{fig:resultse5}, right. A Wilcoxon signed rank test indicated a significant difference ($Z = -3.16, p = 0.0016, Cohen's~d = 1.02$). Note that while there is a 50\% random chance of being correct, this is equal for both conditions. Also, no block-wise differences within conditions or gender-specific differences for error rate were found. 

We further analyzed the error rate according to the left, center and right zones of the displays, with each zone encompassing 4 columns, to check for possible edge effects, i.e. the possibility of a higher error rate in the outer display regions. 
In the condition \textsc{HMD+Projector} the left and right zone correspond to the columns presented on the HMD and the center zone corresponds to the columns presented on the projector. No statistically significant differences between the center and the outer zones could be found, see Figure \ref{fig:results-zones}.

In other words, the error rate using \textsc{HMD+PROJECTOR} significantly increased about 100\% compared to only \textsc{HMD}.

\subsubsection{Workload}

\begin{figure}[tb]
	\centering
	\includegraphics[width=\columnwidth]{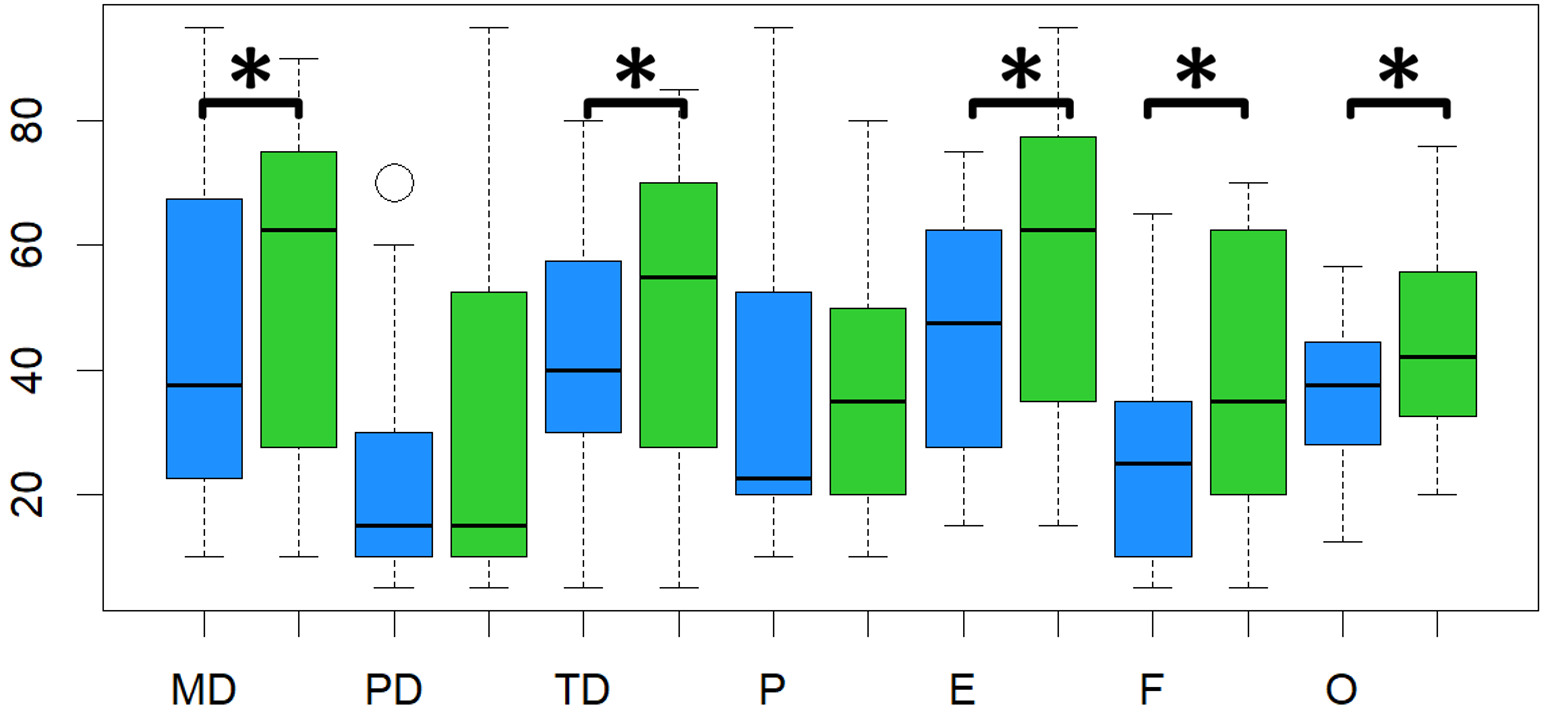}
	\caption{Scores for the NASA TLX questionnaire. From left to right: Mental Demand (MD), Physical Demand (PD), Temporal Demand (TD), Performance (P), Effort (E), Frustration (F) and Overall Demand (O) \textsc{HMD} (blue) and \textsc{HMD+Projector} (green). Significant differences are marked by a $*$ symbol.}	\label{fig:tlx}
\end{figure}

Scores for the workload of both conditions measured by the unweighted NASA TLX questionnaire are depicted in Figure \ref {fig:tlx}.

Wilcoxon signed rank tests indicated significant differences for mental demand ($Z = -2.61, p = 0.009, Cohen's~d = 0.81$), temporal demand ($Z = -2.02, p = 0.043, Cohen's~d = 0.61$), effort ($Z = -3.68, p < 0.001, Cohen's~d = 1.25$), frustration ($Z = -3.09, p = 0.002, Cohen's~d = 0.99$) and overall demand ($Z = -3.44, p = 0.001, Cohen's~d = 1.14$), but not for physical demand or performance. Please note, that a gender-specific difference was indicated for the physical demand rating  in the \textsc{HMD+PROJECTOR} condition between female ($mean~score = 41.67, sd = 28.56$) and male ($mean~score = 17.92, sd = 20.07 $) ($Z = 2.16, p = 0.033, Cohen's~d = 0.66$), but not within each group between \textsc{HMD+PROJECTOR} and \textsc{HMD}.


In other words, the demand when using \textsc{HMD+PROJECTOR} significantly increased compared to only \textsc{HMD} (except for physical demand). Also, pariticpant subjectively did not judge their performance significantly different in both conditions.


\subsubsection{Simulator Sickness} 

The overall results for the simulator sickness questionnaire are depicted in Table 1. Wilcoxon signed rank tests indicated no significant differences for the total SSQ score, nor for the subscales nausea, oculo-motor or disorientation.

\begin{figure*}[t!]
	\centering 
	\includegraphics[width=2\columnwidth]{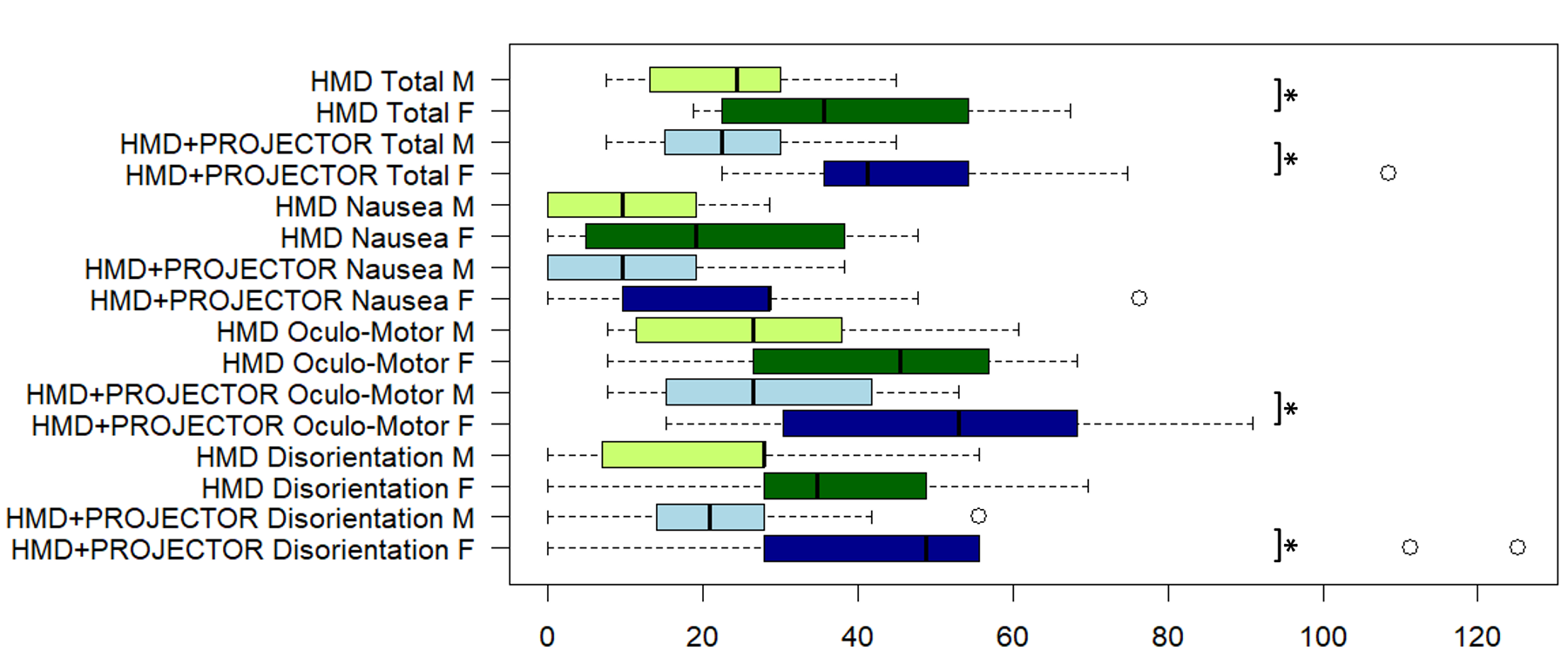}
	\caption{Gender-specific differences of the SSQ questionnaire results. Significant differences are marked by a $*$ symbol. Please note, that no significant difference were found between conditions for male or female participants. F: female. M: male. }
	\label{fig:ssqgender}
\end{figure*}

Please note, that gender-specific differences were indicated for the total score in both \textsc{HMD} ($Z = 2.26, p = 0.025, Cohen's~d = 1.04$) and \textsc{HMD+PROJECTOR} ($Z = 3.13, p = 0.001, Cohen's~d = 1.66$) as well as for oculo-motor ($Z = 2.44, p = 0.015, Cohen's~d = 1.15$) and disorientation ($Z = 2.46, p = 0.015, Cohen's~d = 1.16$) scores in the \textsc{HMD+PROJECTOR} condition between female and male, but not within each group between \textsc{HMD+PROJECTOR} and \textsc{HMD}, see also Figure \ref{fig:ssqgender}.



\begin{table}
\begin{center}
	\begin{tabular}{ |c|c|c| }
		\hline
		Scale & \textsc{HMD} & \textsc{HMD+PROJECTOR} \\
		\hline
		Total & 31.8  & 36.5  \\
		& (16.0) &  (21.9)\\
		\hline
		Nausea & 15.9 & 18.3 \\
		& (15.3) & (18.6) \\
		\hline
		Oculo-Motor & 34.4 & 39.5 \\
		& (19.8) & (21.8) \\
		\hline
		Disorientation & 29.6 & 36.5 \\
		& (20.2) & (30.9) \\
		\hline
	\end{tabular}

\caption{Average SSQ results with standard deviation in parenthesis.}
\end{center}
\label{tab:ssq}
\end{table}

In other words, while gender-specific differences were detected within conditions (as suggested by literature on gender-specific differences in motion sickness \cite{klosterhalfen2005effects, }), no significant differences were detected between conditions for simulator sickness. 

\subsubsection{Search Strategies}
For condition \textsc{HMD}, 10 out of 24 participants mentioned that they saw if the target was present or not at first glance. Nine participants mentioned that they first focused on the image center and, in case they did not spot the target, searched the corners and borders in a clockwise order. The remaining participants conducted the search following an S-shape, starting on the top left of the screen, or divided the screen into four quadrants, which they searched subsequently.

For condition \textsc{HMD+PROJECTOR}, six participants mentioned they always focused on the center of the projection area first, and afterwards on the depth layer of the HMD. Four participants mentioned they always focused first on the background layer and thereafter on the projection in the foreground. Three participants mentioned they exclusively focused on the background layer (such that the foreground was always perceived blurry). The remaining 11 participants indicated they regularly switched between focusing first on the foreground or the background.

No significant differences regarding task completion time, errors, or workload were indicated between those groups.

\section{Discussion and Implications}

The results indicate that conducting a visual search task across two depth layers results in a significantly higher reaction time with an approximate increase of 50\% and significantly higher error rate with an approximately increase of 100\% compared to using a single depth layer. In addition, the perceived workload was higher for two depth layers, however simulator sickness revealed no significant difference.

While previous work has shown that there is an effect of depth layer and context switching \cite{huckauf2010perceptual, gabbard2018effects}, this experiment quantifies that the reaction time and error rate for joint visual information processing across an OST HMD and a body-proximate display are substantially higher compared to visual information processing at a single depth layer. Factors such as display type (e.g., projector vs. smartphone, different OST-HMD models), display parameters (such as resolution) could influence the absolute values of the results, but we expect the magnitute of the results to not substantially change for the difference in viewing distances evaluated in this experiment. However, it would be worthwile to better understand how the performance and subjective parameters change as the vergence and accomodation distances converge.

While the general direction of these results was expected, the extent of the negative effects was previously unknown and highlights the need for thoughtful design when distributing information across an OST HMD and a body-proximate display. As we anticipate such solutions to emerge in the near future as mitigation tactics against the user interface limitations inherent in the current generation OST HMDs in practical applications, the extent of the cost of such solutions is valuable to make informed design decisions.

The results also highlight the need for research to both understand the full extent of the consequences of joint OST HMD and body-proximate display systems and to identify successful mitigation techniques and strategies. While some techniques exists for adaptive sharpening of out-of-focus content, e.g., \cite{oshima2016sharpview, cook2018user}, it remains to be seen, if they can have a positive benefits for joint visual information processing tasks.  

However, the effect of double vision, when verging the eyes between foreground and background layer, remains an open issue. While this could be theoretically addressed by putting the vergence plane of an OST HMD to 30 cm, in practice this would substantially reduce the effective field of view in which stereo rendered content could be perceived. In the future, vari-focal or multi-focal OST HMDs (e.g., \cite{dunn2017wide, akcsit2017near, wilson2018high}) might potentially mitigate the effects indicated in our study.

One option for current generation OST-HMDs might be to minimize the amount of joint information processing across the two depth layers in safety critical situations using such a multi-display setup. This does not implicate to only use a single depth layer all the time, but to carefully direct the attention of users either to the HMD or to the body-proximate display at a time. However, further studies would need to investigate this in detail.

\section{Conclusions and future work}
Joint OST HMD and body-proximate display systems open up a new solution space. However, there are also likely negative effects, such as reduced visual information integration performance among users. To understand the extent of this cost this paper presents the result of an experiment evaluating human performance in a visual search task across an OST HMD and a body-proximate display at 30 cm. 

The results revealed that task completion times significantly increased by approximately 50\% and error rates significantly increased by approximately 100\% compared to a visual search on a single depth layer. 

In addition, perceived workload ratings were higher for joint visual information processing across two depth layers. The results quantify the negative joint visual information integration cost and highlight a design trade-off when exploring joint OST HMD and body-proximate display systems.

In future work, we would like to study the effects of joint information processing acrosss OST HMDs and body-proximate displays in work scenarios, such as following a repair instruction at a physical machine. This would further inform users and manufacturers of current generation HMDs in industrial settings about the expected costs when using HMDs with physical objects within hands reach, specifically as some manufacturers such as Microsoft are pushing OST HMDs as supporting tool for first-line workers \cite{microsoft2019hololens}.

Further, we plan to replicate the study with further OST HMDs such wide field-of-view HMDs (to set the vergence plane at the same distance as the body-proximate display, which would still not resolve the accomodation issue), as well as multi- or varifocal HMDs. Even if the accomodation conflict would be mitigated using those HMDs, research indicates that there still remains a cost for jointly processing the information across an HMD and a further output medium at the same depth layer \cite{huckauf2010perceptual}.

\balance
\bibliographystyle{ACM-Reference-Format}
\bibliography{focus}

\appendix

\end{document}